\documentclass[
	aps,
	prl,
	reprint,
	amsmath,amssymb,
	superscriptaddress,floatfix
]{revtex4-2}
\usepackage{graphicx} 
\usepackage{dcolumn} 
\usepackage{bm}
\usepackage{amssymb,amsmath}
\usepackage{epstopdf}
\usepackage{braket}
\usepackage{color}
\usepackage{xcolor}
\usepackage{mathrsfs}
\usepackage{lipsum}
\usepackage{afterpage}
\usepackage{ragged2e}

\begin{document}
\title{Mechanically Modulated Sideband and Squeezing Effects of Membrane Resonators}
\author{Fan Yang}
\affiliation{Fachbereich Physik, Universit{\"a}t Konstanz, 78457 Konstanz, Germany}
\author{Mengqi Fu}
\affiliation{Fachbereich Physik, Universit{\"a}t Konstanz, 78457 Konstanz, Germany}
\author{Bojan Bosnjak}%
\affiliation{Center for Hybrid Nanostructures, Universit{\"a}t Hamburg, 22761 Hamburg, Germany}
\author{Robert H. Blick}%
\affiliation{Center for Hybrid Nanostructures, Universit{\"a}t Hamburg, 22761 Hamburg, Germany}
\author{Yuxuan Jiang}%
 \email{yuxuan.jiang@ahu.edu.cn}
\affiliation{School of Physics and Optoelectronics Engineering, Anhui University, 230601 Hefei, China}
\author{Elke Scheer}%
 \email{elke.scheer@uni-konstanz.de}
\affiliation{Fachbereich Physik, Universit{\"a}t Konstanz, 78457 Konstanz, Germany}
\
\begin{abstract}
   We investigate the sideband spectra of a driven nonlinear mode with its eigenfrequency being modulated at a low frequency ($<$1 kHz). This additional parametric modulation leads to prominent antiresonance lineshapes in the sideband spectra, which can be controlled through the vibration state of the driven mode. We also establish a direct connection between the antiresonance frequency and the squeezing of thermal fluctuation in the system. Our work not only provides a simple and robust method for squeezing characterization but also opens a new possibility toward sideband applications.
\end{abstract}
\maketitle
\begin{figure}[t!]
   \includegraphics[width=\linewidth]{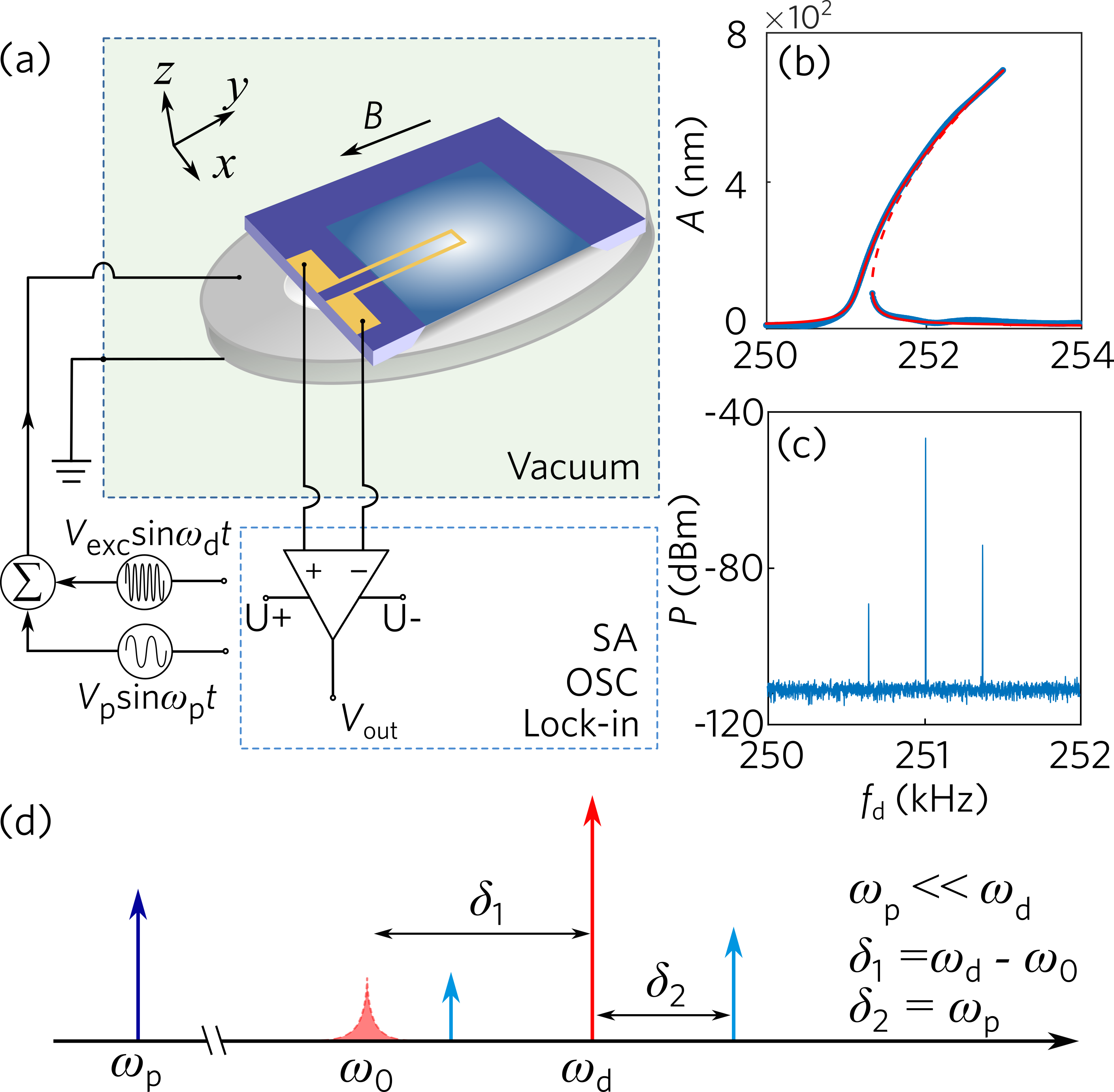}
       \caption{a) Schematic drawing of the silicon nitride (Si-N) membrane resonator structures and the on-chip electromagnetic detection scheme. b) The frequency response curve of the fundamental (1,1) mode (solid blue line) measured at $V_\textrm{exc} = 0.1~\textrm{V}$ fitted by the Duffing model (red dash line). c) The measured power spectrum of a two-tone experiment with $f_\mathrm{d} = 251~\textrm{kHz}$ and $V_\textrm{exc} = 0.5~\textrm{V}$, $f_\mathrm{p} = 380~\textrm{Hz}$ and $V_\mathrm{p} = 0.3~\textrm{V}$. d) Schematic drawing of the measurement scheme and the frequency relation between the drive/probe tone and their response.}
   \label{fig:image01}
\end{figure}
\begin{figure}[t!]
    \includegraphics[width=0.9\linewidth]{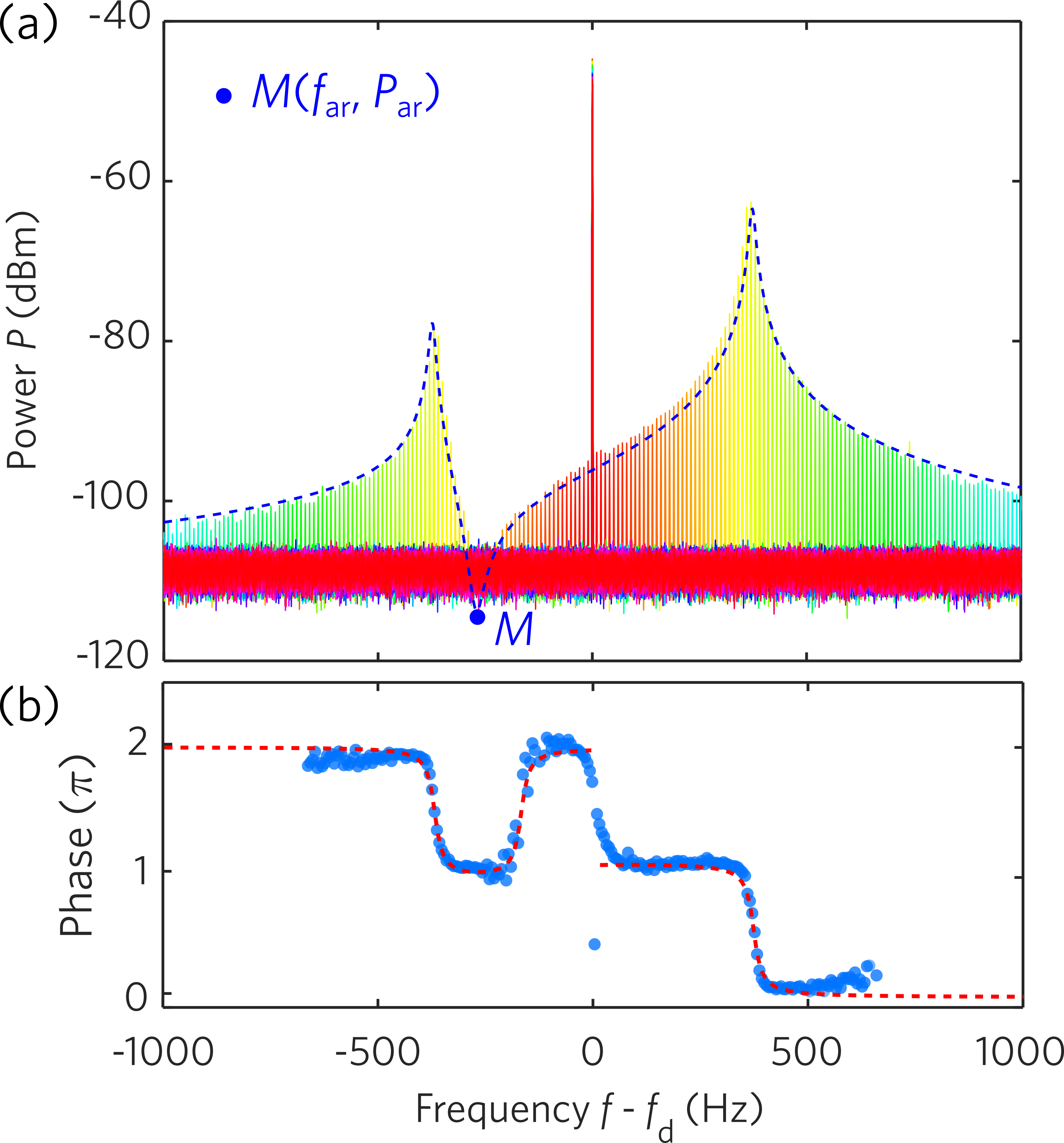}
    \caption{a) Overlaid power spectra with a sweeping probe frequencies $f_\mathrm{p}$ increasing from 10 to 1000 Hz in steps of 10 Hz and $V_\mathrm{p} = 0.5~\textrm{V}$ under a given drive $V_\textrm{exc} = 0.5~\textrm{V}$ and $f_\mathrm{d} = 251~\textrm{kHz}
    $. The color coding indicates different probe frequencies. The blue dashed line shows a parameter free sideband calculation using Eq. \eqref{eq:Mathieu_eq}. The coordinate $M(f_\mathrm{ar},P_\mathrm{ar})$ marks the calculated minimum point of the silent region. b) The phases of the sideband captured under the same condition in a) and the its calculated phase are shown in blue dots and red-dashed line, respectively. Panel a) and b) share the same frequency axis.}
    \label{fig:image02}
\end{figure}

Micro- and nanomechanical resonators have been shown to be ultra sensitive for charge, force and mass measurements in the nonlinear regime \cite{tadokoro2021highly,papariello2016ultrasensitive,li2021novel,zhang2020giant,kim2016ultrananocrystalline,park2013phonon}. However, the high sensitivity also renders the resonators susceptible to environmental fluctuations such as thermal noise \cite{wen2020coherent,cleland2002noise,sansa2016frequency} or molecular
motion \cite{yang2011surface,atalaya2011diffusion}, thereby limiting their applications. One way to circumvent this limit is by taking advantage of squeezing effects, where the fluctuation of one quadrature is reduced at the expense of that in its conjugate \cite{rugar1991mechanical}. Such a signature was firmly established in both theory and experiment to characterize squeezing effects in various types of resonators \cite{wollman2015quantum,lecocq2015quantum,pirkkalainen2015squeezing,dykman2012fluctuating}. However, the squeezing effect in the quadrature is usually very subtle and requires sensitive measurements and careful analysis to enable its detection, particularly in systems of high quality factors ($Q$).\\
\indent Recently, there has been rising interest in the sideband response of  resonantly driven nonlinear modes \cite{huber2020spectral,ochs103amplification,heugel2021ghost}.
In contrast to those arising from the coupling of flexural modes \cite{houri2019limit,ganesan2017phononic,czaplewski2018bifurcation}, these sidebands are related to fluctuations around stable states of the driven system and appear in pairs at either side of the driven frequency. It has been shown that the sidebands exhibit a strong amplitude asymmetry within the pair and their integrated intensity ratios
give direct access to squeezing in the system \cite{huber2020spectral}, which avoids the difficulties in the quadrature method. On the other hand, these sidebands also have important implications in technological applications such as tunable signal-to-noise ratio amplifiers, super-narrow frequency detectors and filters \cite{dykman1994supernarrow}. Therefore, it is of great importance to understand their behavior under external stimulus and achieve the control of these sidebands.\\
\indent In this Letter, we present a so far undescribed sideband response of a nonlinear vibrating membrane resonator using a two-tone measurement. In addition to the drive tone that resonantly excites the system, we also apply an additional probe tone with a very low frequency ($<$1 kHz). We find that this probe tone modulates the eigenfrequency of the system and leads to a sideband response markedly different from previous studies \cite{huber2020spectral,ochs103amplification,heugel2021ghost}. Specifically, both their amplitudes and lineshapes show strong asymmetries and, most saliently, the sidebands exhibit a typical antiresonance response. We further show that the antiresonance frequency can be used to determine the squeezing parameters of the system, and they are tunable through the control of vibrational states of the driven mode. Such a tunable sideband response offers a new possibility for future device applications. Our results may also give insight into other sideband related phenomena such as the energy transfer via mode coupling in the nonlinear and the chaotic regime \cite{heugel2021ghost, yang2019spatial, yang2021persistent, kecskekler2021tuning, houri2020generic}.
\\
\indent The measured device is composed of a $\mathrm{\sim}$ 500 nm thick silicon nitride (Si-N) membrane suspended on a massive silicon frame attached to a piezo ring. The vibration of the freestanding membrane is excited by applying an AC voltage onto the piezo ring at a drive frequency $f_\mathrm{d}$ close to the eigenfrequency $f_0$ of the membrane resonator. To detect the vibrations, we fabricate 27 nm thick aluminum structures on the membrane using a standard e-beam lithography technique and measure the inductive voltage across the structures under an external uniform magnetic field, as depicted in Fig. \ref{fig:image01}a). The driven mode parameters (frequency, amplitude, quality factor) can be quantitatively obtained from the voltage using the Faraday's law. All the measurements are performed in vacuum at room temperature. More sample fabrication and setup details can be found in the Supplementary Materials (SM) \cite{SM} and also our previous works \cite{yang2017quantitative,waitz2012mode,waitz2015spatially}.\\
%
%
%
%
\indent Figure \ref{fig:image01}b) shows the response of our system to a typical single excitation $V_\textrm{exc}(t) = V_\textrm{exc}\sin(2\pi f_\mathrm{d} t)$ driven close to its fundamental mode. The vibration exhibits a typical Duffing response with two stable states of different vibration amplitudes when the drive tone is sweeping along different direction, and we can quantitatively extract the driven mode parameters including its eigenfrequency $f_0 \approx 250.85~\mathrm{kHz}$, damping $\Gamma/2\pi = 13.5~\mathrm{Hz}$ ($Q \approx 18000$) and the cubic (Duffing) nonlinearity $\beta = 1.13 \times 10^{23}~\mathrm{m}^{-2}\mathrm{s}^{-2}$ using the methods described in the SM \cite{SM}. In the upper branch (i.e., the large vibration state), a simultaneous application of a drive tone at $f_\mathrm{d}$ and a probe tone with a much smaller frequency $f_\mathrm{p}$ can result in the excitation of two ultranarrow sidepeaks with their frequency at $f_\mathrm{d}+f_\mathrm{p}$ (blue sidepeak) and $f_\mathrm{d}-f_\mathrm{p}$ (red sidepeak), respectively, as indicated by our experiments in Fig. \ref{fig:image01}c). These frequency relations are schematically summarized in Fig. \ref{fig:image01}d). From here, we can safely exclude the possibility of a sidepeak induced by cubic nonlinearity, where the sidepeak should locate at two times the probe frequency away from the drive frequency \cite{reichenbach2005third}. \\
\indent We then sweep the probe tone from 10 to 1000 Hz with a step size of 10 Hz under a fixed drive tone, and the sideband response is plotted in Fig. \ref{fig:image02}a). Their amplitude reaches a maximum at a peak frequency  $f_\mathrm{pk} = \pm~380~\mathrm{Hz}$ and starts to decrease if $f_\mathrm{p}$ further increases. More importantly, there is a prominent silent region where the sideband signals are strongly suppressed below the noise floor. This region only appears in the red sideband but is absent in the blue sideband, leading to a drastically asymmetric lineshape. We will show below that there exists a minimum point $M$ in the silent region with its frequency and amplitude labeled by $f_\mathrm{ar}$ and $P_\mathrm{ar}$, respectively. The sidebands also exhibit interesting phase response as depicted by the blue dots in Fig. \ref{fig:image02}b). There are phase shifts of $\pi$ when the ultranarrow sidepeaks are tuned through the drive tone $f_\mathrm{d}$, the two sideband peaks at $f_\mathrm{d} \pm f_\mathrm{pk}$, and the $M$ point at $f_\mathrm{d}+f_\mathrm{ar}$. All these features in both the amplitude and phase are reminiscent of a typical antiresonance response \cite{sames2014antiresonance,wahl1999significance}. More sideband spectra measured under different drive conditions are shown in the SM \cite{SM} and they all exhibit similar antiresonance response.\\
%
%
%
%
\indent To model the observed sideband response, we use the following equation of motion:
\begin{equation}
    \ddot{x}+2\Gamma \Dot{x}+\left(\omega_0^2-a^2\cos(\omega_\mathrm{p}t)\right)x+\beta x^3=F\cos(\omega_\mathrm{d}t),
    \label{eq:Mathieu_eq}
\end{equation}
where $F,\,\omega_{0} = 2\pi f_0,\,\omega_\mathrm{d}= 2 \pi f_\mathrm{d}$, and $\omega_\mathrm{p}= 2 \pi f_\mathrm{p}$ denote the driving force, the angular frequency of the flexural mode, the drive tone and probe tone, respectively. The probe tone also modulates the eigenfrequency with a strength of $a^2$ which can be characterized via a DC measurement with the method in the SM\cite{SM}.\\
\indent Using the standard rotating wave approximation, we can switch from the lab frame into the rotating frame with two new quadratures $y$ and $y^*$. We can approximate the solution by expanding the quadrature $y$ into a main response $y_0$ and the perturbed response $y_1$. The former describes the response at $\omega_\mathrm{d}$ while the latter is responsible for the sidebands. We find that $y_0$ is governed by the well-known Duffing equation, and the equation of motion for $y_1$ is 
\begin{equation} \dot{y_1}=-\Gamma(1+i\delta \omega-2i|y_0|^2)y_1+i\Gamma y_0^2 y^*_1
-i\Gamma \cos (\omega_\mathrm{p}t)y_0,
\label{order1m}
\end{equation} 
with $\delta \omega=(\omega_\mathrm{d}-\omega_0)/\Gamma$. Here, the first two terms on the right hand side of the equation describe the vibrations when the resonator deviates from the stable states and the last term represents a force induced by the parametric modulation. By solving this equation, we obtain the sideband amplitudes as well as their phases at different probe frequencies. Further details of the theoretical calculation can be found in the SM \cite{SM}. \\

\indent In Fig. \ref{fig:image02} a) and b), we compare the theoretical calculations (dashed lines) with experiment data of the amplitude and phase of the sideband response, respectively and they both showed excellent agreements. We note that the $\pi$ phase shift across the drive frequency is not an intrinsic effect but due to a phase delay between the longitudinal and transverse waves in the piezo \cite{cleland2013foundations}. Further discussion can be found in the SM \cite{SM}.\\
\indent From here, we can clearly see that the silent region in experiment corresponds to an antiresonance dip $M$ in the sideband response. In fact, the emergence of a clear antiresonance dip requires $\omega_\mathrm{pk}/\Gamma \gg 1$, and in this weak damping limit 
\footnote{
Note that here the weak damping limit adopt the requirement of squeezing parameter in the following discussion for simplification.}
, we find that $f_\mathrm{ar}$ can be well approximated by
\begin{equation}
    \omega_\mathrm{ar}=\Gamma \sqrt{1+(\delta\omega-|y_0|^2)^2}.
    \label{eq:dip_freq1}
\end{equation}
This relation suggests that the position of the $M$ point is susceptible to the vibration state of the driven mode, as $y_0$ is determined by both the drive and the detuning of the drive tone (see SM \cite{SM}). 

\begin{figure}[t!]
 \includegraphics[width=\linewidth]{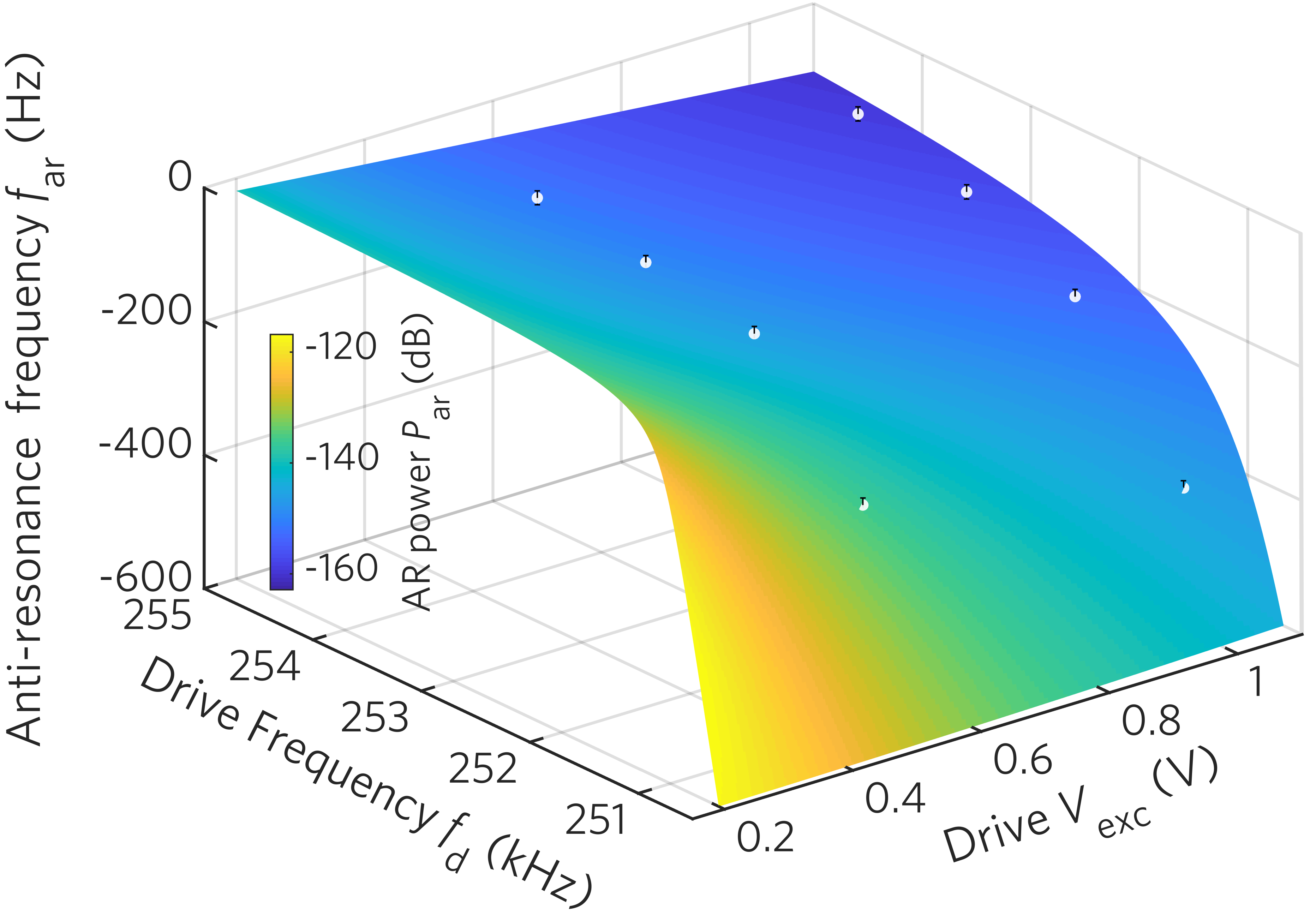}
     \caption{Theoretically calculated antiresonance frequencies $f_\textrm{ar}$ and their power $P_\mathrm{ar}$ under different drive frequency $f_\mathrm{d}$ and power $V_\textrm{exc}$. The white spheres are the experimentally extracted $f_\mathrm{ar}$ at the corresponding condition with an error bar of $\pm 10$ Hz.} 
 \label{fig:image03}
\end{figure}

\begin{figure}[t]
    \includegraphics[width=\linewidth]{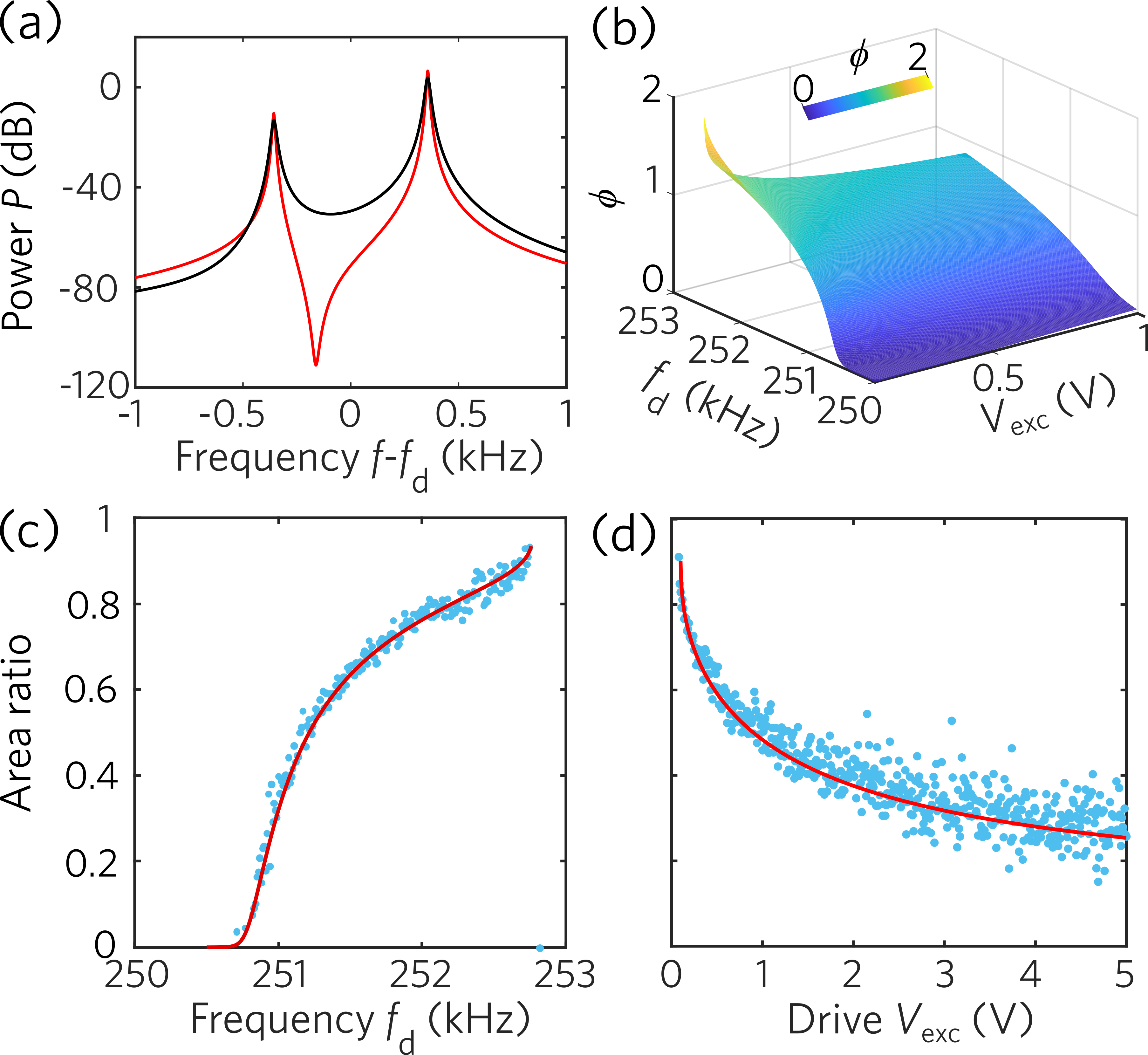}
    \caption{a) Theoretical sideband amplitudes driven by broadband noise (black) and a probe tone (red) with the same drive tone. b) Calculated squeezing parameter $\phi$ using $f_\textrm{ar}$ under different drive condition. c) The detuning dependence of the intensity area ratio between the left and the right sideband $I_{-}/I_{+}$ under $V_\textrm{exc}$ = 0.1 V. d) The drive power dependence of $I_{-}/I_{+}$ measured at $f_\mathrm{d}$ = 252.5 kHz. In both c) and d), the noise bandwidth is 300 kHz and $V_\textrm{noise}$ = 0.4 V. The theoretical results from our model are shown as red solid lines and the experiments are shown as blue dots, and they share the same y axis.}
    \label{fig:image04}
\end{figure}

Figure \ref{fig:image03} shows the calculated $f_\mathrm{ar}$ and $P_\mathrm{ar}$ as a function of power and detuning of the drive tone using the mode parameters given above. To compare with experiments, we can extract $f_\mathrm{ar}$ using the center frequency of the silent region, and the error bars correspond to the width of the silent region. These experimentally extracted positions (white sphere symbols) agree well with our theoretical calculation, as indicated in Fig. \ref{fig:image03} and we can also make the following observations: (1) The highest power in the calculations here is still 10 dBm less than our experimental noise level, indicating an excellent cancellation effect of the sideband motion; (2) It is also evident here that $f_\mathrm{ar}$ has a lower bound given by the damping $\Gamma/2\pi$, as suggested from Eq. \eqref{eq:dip_freq1}; (3) When the power of the drive tone is large or its detuning is small, $f_\mathrm{ar}$ is significantly away from the drive frequency.
\\
%
%
%
%
\indent Further insights can be gained by comparing our results with the sidebands induced by thermal fluctuations \cite{huber2020spectral}. In Fig. \ref{fig:image04}a), we compare a theoretical sideband spectrum of our case (red line) with that by thermal fluctuation (black line), both using the mode parameters from our experiments. Except for the antiresonance feature, the two spectra have very similar lineshapes, particularly regarding the peaks. Therefore, both types of sidebands reflect the vibration modes around the stable states of the driven mode (i.e., sideband). The peak positions of the sidebands correspond to the eigenfrequency of these vibration modes and can be determined by
\begin{equation}
    \omega_\mathrm{pk}=\Gamma\cdot \mathbf{Im}\left( \sqrt{(3|y_0|^2-\delta \omega)(|y_0|^2-\delta \omega)}\right),
    \label{eq:peak_freq}
\end{equation}
where $\mathbf{Im}$ indicates the imaginary part of the argument. The emergence of an antiresonance in the sideband spectrum reveals the different character of the drive forces: a deterministic coherent single-frequency force that periodically changes the system parameters (appearing as combined amplitude-frequency modulation here, see SM \cite{SM}) destructively interfere with sideband while this effect is absent in the sidebands induced by broadband stochastic noise. Based on this observation, these sidebands can be delineated as ``quasi-modes'' characterized as eigenstates controlled by the vibration states of the mechanical system in the nonlinear regime. \\
\indent Even though the sidebands induced by parametric modulation is a deterministic one, we can nevertheless relate it with the squeeezing effect of the fluctuations in the system. In the weak damping limit, there is a simple connection between the squeezing parameter $\phi$ that measures the transfer of fluctuations between a pair of conjugate variables and the antiresonance frequency in the spectrum:
\begin{align}
\label{eq:squeezing1}
    \tanh{2\phi}=
\frac{\omega_\mathrm{ar}+\Gamma \delta\omega}{2\omega_\mathrm{ar}+\Gamma \delta\omega}.
\end{align}
We note that the above relation only applies when the system is in the upper branch. In principle, a similar expression for the lower branch can be derived. However, in experiment, it is difficult to identify an antiresonance structure in sideband spectrum due to its low amplitude. (see SM \cite{SM}).\\
%
%
%
%
\indent To testify the above relations, we compare the squeezing parameters extracted from antiresonance frequency and those from the noise induced sideband area ratio. For a thermally-induced sideband, it has been shown that the integrated intensity ratio between the two sidebands ($I_{-}/I_{+}$) is solely determined by the squeezing parameter $\phi$ \cite{huber2020spectral}. Due to a moderate $Q$ in our device, the environmental thermal fluctuation itself does not induce detectable sidebands. Nevertheless, we can still access this quantity by introducing a broadband white noise of 300 kHz bandwidth and $V_\textrm{noise}$ = 0.4 V \cite{SM} in addition to a drive tone of $V_\textrm{exc}$ = 0.1 V and $f_\mathrm{d}$ = 252.5 kHz. The integrated intensity can be obtained by fitting the sidebands around the peak frequency with two independent Lorentzian functions. Fig. \ref{fig:image04}c) and d) show the detuning and drive power dependence of the integrated intensity ratio in blue dots while the red lines in Fig. \ref{fig:image04} c,d) denote the theoretical calculated values using $f_\textrm{ar}$, Eq. \eqref{eq:dip_freq1} and Eq. \eqref{eq:squeezing1}. We emphasize that there is no free parameter used in these calculations and the excellent agreement between the two methods indicates that $f_\textrm{ar}$ is indeed a reliable indicator of the squeezing parameter. \\
\indent Finally, we wish to comment on the robustness and simplicity of our method. In the previously reported method\cite{huber2020spectral}, it requires a separate fitting for every sideband spectrum to extract the squeezing parameter. In contrast, our antiresonance frequency method only relies on the knowledge of three parameters: $\delta \omega$, $\Gamma$, and $\phi$, all of which can either be directly extracted from the experiment or from a one-time fitting with the driven modes, making our method much easier and simpler. The antiresonance frequency also provides a direct way to compare the squeezing parameters under different conditions. Figure \ref{fig:image04}b) shows the calculated $\phi$ over the same parameter space as that in Fig. \ref{fig:image03}, and it is positively related to both the detuning and force. By comparing this result to the $f_\textrm{ar}$ map in Fig. \ref{fig:image03}, we can deduce that the distance between $f_\textrm{ar}$ and the dirve frequency is inversely related to $\phi$: the smaller the distance, the larger $\phi$. 
\\
\indent In conclusion, we measured the sideband response of a free-standing Si-N membrane using a two-tone measurement in its nonlinear regime with the drive frequency much larger than that of the probe tone. We observed a typical antiresonance lineshapes in the sideband spectra. The amplitude and the phase of these sidebands can be well described by a Duffing equation with its eigenfrequency being parametrically modulated at the probe-tone frequency. The antiresonance results from a destructive interference between the probe tone and the resulting sideband. We further demonstrate that the antiresonance can be controlled by the vibrational state of the driven mode, thus providing a new possibility for future application of sidebands. For example, these sidebands can act as ``quasi-modes'' and couple with other vibrational modes for energy reshuffling in a nonlinear system. Finally, we provide a robust and simple method to 
characterize the squeezing effect even for resonators with a moderate $Q$, evidenced by the excellent agreement between the squeezing parameter extracted with the antiresonance frequency and those with the sideband-area method.\\
\indent The authors thank J. Boneberg for help with the experimental setups. We are indebted to J. S. Ochs, E. M. Weig, G. Rastelli, and W. Belzig for fruitful discussion and comments about the work. The authors gratefully acknowledge financial support from the China Scholarship Council, the A. v. Humboldt Foundation, the Deutsche Forschungsgemeinschaft  (DFG, German Research Foundation) through SFB 1432 (Project-ID 425217212), and the Wisconsin Alumni Research Foundation (WARF) via the Accelerator Program.\\

\indent F. Yang. and M. Fu contributed equally to this work. 
\bibliographystyle{apsrev4-1}
\bibliography{modulation_references}
\end{document}


\title{Supplemental Material for \\Mechanically Modulated Sideband and Squeezing Effects of Membrane Resonators}
\author{Fan Yang}
\affiliation{Fachbereich Physik, Universit{\"a}t Konstanz, 78457 Konstanz, Germany}
\author{Mengqi Fu}
\affiliation{Fachbereich Physik, Universit{\"a}t Konstanz, 78457 Konstanz, Germany}
\author{Bojan Bosnjak}%
\affiliation{Center for Hybrid Nanostructures, Universit{\"a}t Hamburg, 22761 Hamburg, Germany}
\author{Robert H. Blick}%
\affiliation{Center for Hybrid Nanostructures, Universit{\"a}t Hamburg, 22761 Hamburg, Germany}
\author{Yuxuan Jiang}%
 \email{yuxuan.jiang@ahu.edu.cn}
\affiliation{School of Physics and Optoelectronics Engineering, Anhui University, 230601 Hefei, China}
\author{Elke Scheer}%
 \email{elke.scheer@uni-konstanz.de}
\affiliation{Fachbereich Physik, Universit{\"a}t Konstanz, 78457 Konstanz, Germany}

\maketitle
%
%
\section{Theory: Parametric modulatied Duffing model}

In this section, we present the details of the theory for the antiresonance effect in sidebands. It is known that the frequency response of a rectangular membrane can be well described by the Duffing equation \cite{yang2019spatial}. In our experiment, the membrane is also subject to an additional probe tone whose frequency is much less than that of the drive tone. From Section IV in SM, we find that the probe tone periodically modulates the eigenfrequency of the system. Therefore, the equation of motion for the membrane amplitude $x$ is
\begin{equation}
    \ddot{x}+2\Gamma \Dot{x}+\left(\omega_0^2-a^2\cos(\omega_\mathrm{p}t)\right)x+\gamma x^3=F\cos(\omega_\mathrm{d}t).
\end{equation}
Here, $\Gamma$ is the damping rate, $\omega_0$ is the frequency of an eigenmode, and $\gamma$ is the Duffing nonlinearity. In addition, $a^2$ describes the eigenfrequency modulation strength from the probe tone, and $F$ describes the drive tone with their angular frequency at $\omega_p$ and $\omega_d$, respectively. This is the well known parametric modulation Duffing equation, only that our modulation frequency does not satisfy those commonly used parametric relation such as $\omega_d/\omega_p=2,3$ etc..

It is convenient to solve the equations in the rotating frame for $\omega_d$ via the following transformations:
\begin{align*}
    x=\sqrt{\frac{2\omega_\mathrm{d}\Gamma}{3\gamma}}\left[y(t)e^{i\omega_\mathrm{d}t}+y^*(t)e^{-i\omega_\mathrm{d}t}\right], \qquad
\dot{x}=i\omega_\mathrm{d}\sqrt{\frac{2\omega_\mathrm{d}\Gamma}{3\gamma}}\left[y(t)e^{i\omega_\mathrm{d}t}-y^*(t)e^{-i\omega_\mathrm{d}t}\right],
\end{align*} 
where the $*$ symbol indicates complex conjugates and $i$ is the imaginary unit. Now the equation of motion for the new coordinate in the rotating frame $y$ reads:
\begin{align}
\label{EOM}
\dot{y}=-\Gamma \left[1+i\delta\omega-i|y|^2\right]y -i\Gamma\beta-i\epsilon\Gamma  \cos(\omega_\mathrm{p}t)y,
\end{align}
and the dimensionless parameters are 
$$\delta \omega=\frac{\omega_\mathrm{d}-\omega_0}{\Gamma},\qquad \epsilon=\frac{a^2}{2\omega_\mathrm{d}\Gamma},\qquad \beta=F\sqrt{\frac{3\gamma}{32\omega^3 \Gamma^3}}.$$
Note that in our experimental system we have $\omega_\mathrm{d}\sim\omega_0 $, and $\epsilon$ is a small quantity.

 Since Eq. \eqref{EOM} is a Duffing type equation with the last term being a small correction, we can use perturbation theory and expand the solution $y$ to first order of $\epsilon$:
$$y=y_0+\epsilon y_1+O(\epsilon^2).$$ Substituting the above expansion into Eq.\eqref{EOM} and collecting terms of the same order, we have
\begin{align}
\label{order0}
&\epsilon^0:\quad \dot{y_0}=-\Gamma(1+i\delta \omega-i|y_0|^2)y_0-i\Gamma\beta
\end{align}
\begin{equation}
\begin{aligned}
\epsilon^1:\quad \dot{y_1}=&-\Gamma(1+i\delta \omega-2i|y_0|^2)y_1+i\Gamma y_0^2 y^*_1-i\Gamma\cos(\omega_\mathrm{p}t)y_0.
\end{aligned}
\label{order1}
\end{equation}
The zeroth order ($\epsilon^0$) solution gives exactly the Duffing equation
$$
y_0=\frac{-i\beta}{1+i\delta \omega-i|y_0|^2}.
$$
The first order correction ($\epsilon^1$) describes the sideband motion arising from vibrations around stable states \cite{huber2020spectral} with a complex drive. Since the frequency response at $\omega_\mathrm{p}$ (equivalently $\omega_d \pm \omega_\mathrm{p}$ in the lab frame) is of concern, we assume:
\begin{align}
\label{yassump}
y_1=A_+e^{i\omega_\mathrm{p} t}+A_-e^{-i\omega_\mathrm{p} t}.
\end{align}
Substituting Eq.\eqref{yassump} into Eq. \eqref{order1}, we get
\begin{align}
\left[(1+i\delta\omega-2i|y_0|^2)+i(\omega_\mathrm{p}/\Gamma)\right]A_+=i y_0^2 A^*_{-}-i \frac{y_0}{2}\\
\left[(1+i\delta\omega-2i|y_0|^2)-i(\omega_\mathrm{p}/\Gamma)\right]A_{-}=i y_0^2 A^*_{+}-i \frac{y_0}{2}.
\end{align}
The above equations determines the complex amplitudes
\begin{equation}
\begin{aligned}
\label{eq:modulate}
A_+(\omega_\mathrm{p})=A_{-}(-\omega_\mathrm{p})=-\frac{y_0}{2}\frac{i+\delta\omega-|y_0|^2-\omega_\mathrm{p}/\Gamma}{1+(\delta\omega-2|y_0|^2)^2-|y_0|^4-(\omega_\mathrm{p}/\Gamma)^2+(2i\omega_\mathrm{p}/\Gamma)},
\end{aligned}
\end{equation}

From these expression, we can extract both the phase and the amplitude of the sideband vibrations. We note that if one takes into account a phase delay between the drive and probe tone i.e. $\cos(\omega_p t+\psi)$, it would generate a different phase shift between the blue and red sidebands, which follows 
\begin{align}
\label{phase}
e^{i\psi}A_+=e^{-i\psi}A_-.  
\end{align}

To find out the antiresonance frequency position $\omega_\mathrm{ar}$, one can solve for the frequency that sets the first derivative of the vibration amplitude zero. However, this approach is not easy as it requires solving a fifth order polynomial equations. 
Alternatively, we can define a contrast function $g(\omega_\mathrm{p})$ that gives the magnitude ratio at $\pm \omega_\mathrm{p}$, $$g(\omega_\mathrm{p})=\frac{|A_+(\omega_\mathrm{p})|^2}{|A_{-}(\omega_\mathrm{p})|^2}=\frac{1+(\delta\omega-|y_0|^2-\omega_\mathrm{p}/\Gamma)^2}{1+(\delta\omega-|y_0|^2+\omega_\mathrm{p}/\Gamma)^2}$$
and $\omega_\mathrm{ar}$ can be approximated by the solution that satisfy $dg(\omega_p)/d\omega_\mathrm{p}=0$. With this approach, we find $\omega_\mathrm{ar}$ as  \begin{equation}
    \omega_\mathrm{ar}=\Gamma \sqrt{1+(\delta\omega-|y_0|^2)^2}.
    \label{eq:dip_freq}
\end{equation} and the dip power is $P_\mathrm{ar}=|A_+(\omega_\mathrm{ar})|$. 
We note that the contrast function approach is accurate when the dip is very sharp. Roughly speaking, this corresponds to the weak damping limit (i.e., $\omega_\mathrm{pk}/\Gamma \gg 1$).

\indent To connect $\omega_\mathrm{ar}$ to the squeezing parameter $\phi$, we note that in weak damping limit, the squeezing parameter can be calculated through \cite{huber2020spectral},
$$
\tanh{2\phi}=\frac{|y_0|^2}{2|y_0|^2-\delta \omega}.
$$
Since $|y_0|^2-\delta \omega$ is positive (negative) for the upper (lower) branch, we have
\begin{align}
\label{eq:squeezing}
    \tanh{2\phi_{\pm}}= 
\frac{\omega_\mathrm{ar} \pm \Gamma \delta\omega}{2\omega_\mathrm{ar} \pm \Gamma \delta\omega}.
\end{align}
Here, the $\pm$ corresponds to the high and low amplitude state of the system, respectively. 

\section{Sample fabrication and experiment methodology}
\begin{figure}[htbp]
    \includegraphics[width=0.7\linewidth]{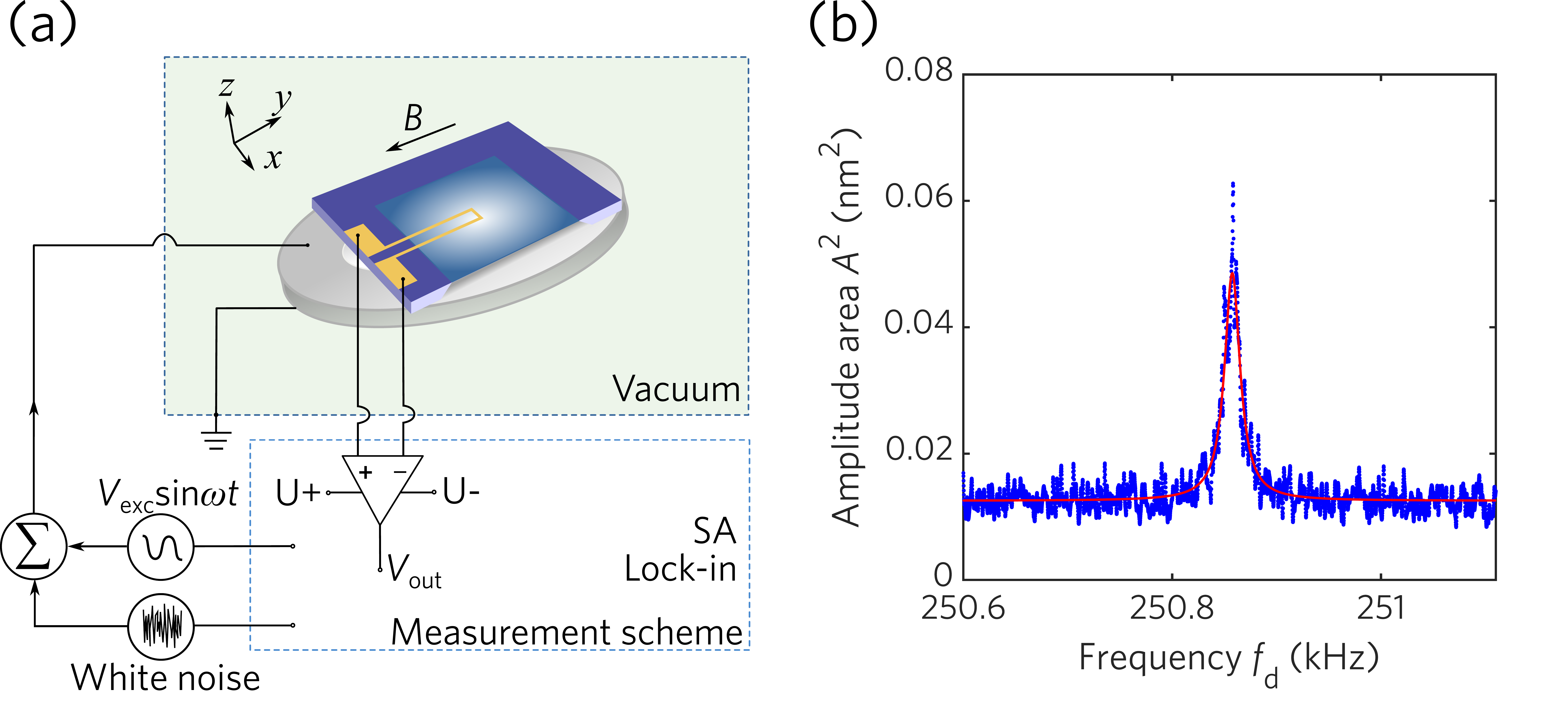}
    \caption{The on-chip nanoelectrode detection scheme on a silicon nitride (Si-N) membrane resonator and the measurement scheme. a) The sketch describes the sideband measurement scheme using a drive tone $V_\textrm{exc} \cdot \sin 2\pi f_\mathrm{d} t$ and and a noise signal $V_\textrm{noise}$ applied to the piezo actuator and measuring the response with a spectrum analyzer. b) The noise-driven amplitude response of the (1,1) mode (blue), fitted by a Lorentzian function and plotted as red solid line ((bandwith of the noise $f_\textrm{BW}~=~300~\textrm{kHz}$, amplitude of the noise $V_\textrm{noise}~=~0.4~\textrm{V}$)).}
    \label{fig:SM_sketch_noise}
\end{figure}
The  Si-N membranes are fabricated from a 0.5 mm thick commercial (100) silicon wafer. Both sides of the silicon substrate are coated with $\sim$ 500 nm thick low-pressure chemical vapor deposited (LPCVD) Si-N. The membrane is fabricated on the front layer. The backside layer serves as an etch mask. Laser ablation is used to open the etch mask with a typical size of 1.5 $\mathrm{\times}$ 1.5 mm$^{2}$. Using anisotropic etching in aqueous potassium hydroxide (KOH), a hole is etched through the openings of the mask. After the KOH solution reaches the topside layer, a membrane is formed, supported by a massive silicon frame. In the present work, the membrane is  $\sim$ 500 nm thick and 542 $\mathrm{\times}$ 524 $\mu$m$^{2}$ in lateral size. The chip with about $1 \times 1 $ $\textrm{cm}^2$ lateral size carrying the membrane is glued to a piezo ring of 20 mm diameter and 5 mm thickness using a two-component adhesive .\\
\indent In this work, we use the magnetic induction method to characterize the amplitude of the membrane. As depicted in both Fig. 1(a) in the main text and Fig. \ref{fig:SM_sketch_noise}a), A $\sim$ 27 nm thick Al electrode is deposited onto the membrane with a detection lead of $L=30~\mu$m long located at the position of interests. Then, the whole device is placed in the vacuum chamber with a pressure of $P~=~10^{-6}~\textrm{mbar}$ at room temperature and subject to an in-plane magnetic field $B$ perpendicular to the detection lead. When the membrane is driven by the piezo with the drive voltage of $V_\textrm{exc}\cdot\sin \omega_\mathrm{d} t$ and the frequency of $\omega_\mathrm{d}=2\pi f_\mathrm{d}$, the magnetic flux through the area enclosed by the detecting electrode and the leads changes with the membrane vibration. Thus, an AC voltage signal is generated in the detecting electrodes and can be detected at the output port of the leads. Here, the leads parallel to the magnetic fields do not contribute to the output voltage ($V_\textrm{out}$) signal and hence the deflection amplitude at the position of the detecting electrodes can be calculated by
\begin{equation}
    A = \frac{V_\textrm{out}}{BL*\omega_\mathrm{d}},
    \label{eq:nm_conver}
\end{equation}
where $A$ is the amplitude, $B=0.45 T$ , and $V_\textrm{out}$ is the voltage before the differential amplifier. 
As an example, the noise-driven spectrum response of the (1,1) mode is shown in Fig. \ref{fig:SM_sketch_noise} b), the Lorentzian fitting curve is plotted as red solid line. From the fitting, the full width at half maximum (FWHM) of the noise-driven response of the (1,1) mode is $\Gamma/2\pi\simeq 14~\textrm{Hz}$, similar to the fitting results from the frequency response of the (1,1) mode.
\section{Sample characterization}
\subsection{Mechanical properties characterization} \label{sec:Mech_properties}
We first summarize the mode parameters used in our experiments and the mechanical properties of the membrane characterized. The Si-N membrane has a lateral size of 524 $\mu \textrm{m} ~\times $ 545 $\mu \textrm{m}$ and its thickness is $\mathrm{\sim} 500~\textrm{nm}$. For the (1,1) mode, the eigenfrequency $f_0$ is around 250.8 kHz; slight shifts may occur due to the drift in the environmental temperature \cite{yang2017quantitative}. The damping is fitted to be $\Gamma/2\pi$ = 13.5 Hz for both the linear and the Duffing regime, giving a quality factor $Q \approx 18000$. The nonlinearity of the (1,1) mode is fitted to be $\beta = 1.13 \times 10^{23}~\textrm{m}^{-2}\textrm{s}^{-2}$. The conversion factor between the voltage and the corresponding force/mass is $5.52\times 10^{-4}$N/kg V, and the typical drive voltage used in this work is 0.1V and 0.5V.
\subsection{ Eigenfrequency modulation characterization}
\begin{figure*}[htbp]
    \includegraphics[width=0.7\linewidth]{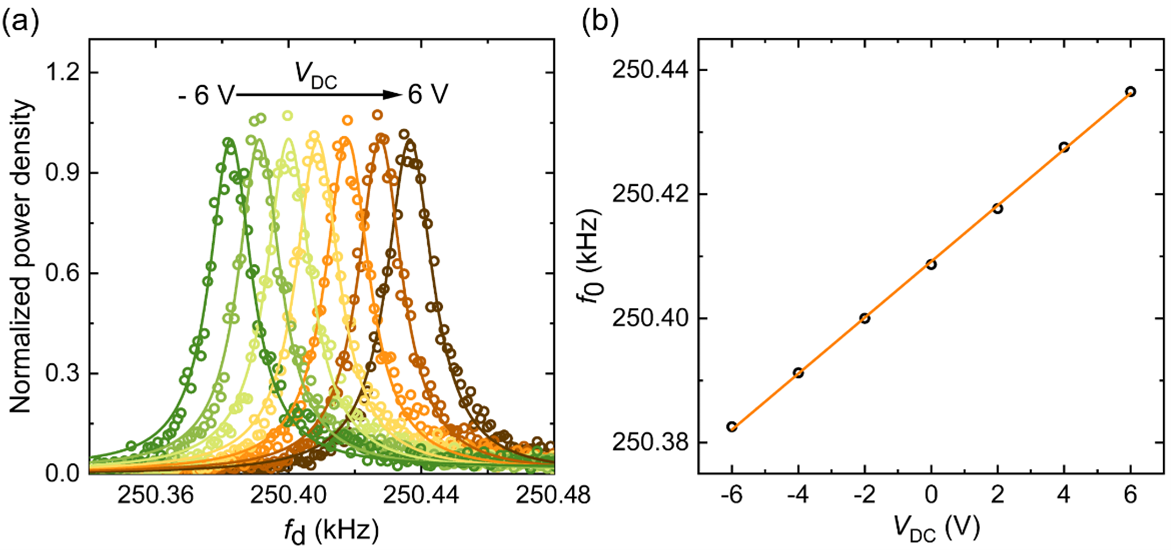}
    \caption{Mechanical tuning characterization. a) The frequency response of the (1,1) mode of the membrane resonator at various $V_\textrm{DC}$ (DC voltage offsets), here $V_\textrm{exc}$ is set to 2 mV for all curves. b) The shift of the eigenfrequency $f_0$ of the (1,1) mode as a function of the $V_\textrm{DC}$.}
    \label{fig:SM_stiff_mod}
\end{figure*}

Figure \ref{fig:SM_stiff_mod}a) presents the frequency response (dots) of the (1,1) mode and their Lorentzian fitting curves (solid lines) at a fixed  $V_\textrm{exc}$ = 2 mV and various DC voltage ($V_\textrm{DC}$) applied to the piezo. Under these condition, the membrane resonator is in the linear regime with similar $Q$ factor at different $V_\textrm{DC}$. However, the eigenfrequency $f_0$ shifts to higher values as $V_\textrm{DC}$ increases. Fig. \ref{fig:SM_stiff_mod}b) shows the dependence between the $f_0$ values extracted by the Lorentzian fitting (dots) and $V_\textrm{DC}$, and the solid line indicates the best linear fit with a slope of  $\Delta f$ = + 4.5 Hz/V. Similar dependence has been observed in other vibration modes.

According to the equation of motion for an driven linear resonator:
\begin{equation}
    \ddot{x}+ \frac{\omega_0^\prime}{Q}\dot{x}+ \frac{k_\textrm{eff}}{m}x = \frac{F}{m}\cos(\omega_\mathrm{d}t).
\end{equation}
 The eigenfrequency $f_0 = \omega_0 /2\pi$ is determined by the effective stiffness of the resonator, $k_\textrm{eff} = m{\omega_0^\prime}^2$. 
 The change of $f_0$ at various $V_\textrm{DC}$ indicates that $k_\textrm{eff}$ can be controlled by an additionally applied voltage on the drive system.
In our two-tone measurement, when a low-frequency probe tone $\omega_\mathrm{p}$ is introduced, $k_\textrm{eff}$ oscillates with the low-frequency excitation as:
\begin{equation}
    k_\textrm{eff} = m[\omega_0 + V_\mathrm{p}\cdot \Delta \omega \cdot \cos(\omega_\mathrm{p}t)]^2
\end{equation}
Here we neglect higher orders in $\cos(\omega_\mathrm{p}t)$, since the frequency modulation in our experiment is much smaller than the eigenmode frequency. The effective stiffness of the vibrating mechanical system can be approximated as:
\begin{equation}
\begin{aligned}
    k_\textrm{eff} \approx k_0 + 2\omega_0\cdot V_\mathrm{p}\cdot\Delta \omega \cdot \cos(\omega_\mathrm{p}t).
\end{aligned}   
\end{equation}
Thus, we can obtained the stiffness modulation parameter $a^2 = 2\omega_0\cdot V_\mathrm{p}\cdot\Delta \omega$, here $\Delta \omega = 2\pi\Delta f$.\\

\indent The origin of the stiffness tuning is due to the transverse component of the Rayleigh wave of the piezo actuator, which modulates the tensile stress of the membrane and hence the $\omega_0$. However, the transverse Rayleigh wave is known to have a $\pi/2$ delay from the longitudinal excitation $\cos(\omega_\mathrm{p}t)$ we send into the piezo \cite{cleland2013foundations}. According to Eq.\eqref{phase}, we would expect an overall $\pi$ phase shift across the drive tone as shown in the Fig.2 b) in the main text.
\section{Additional experimental data on the spectral response of sideband}
\begin{figure*}[htbp]
    \includegraphics[width=0.7\linewidth]{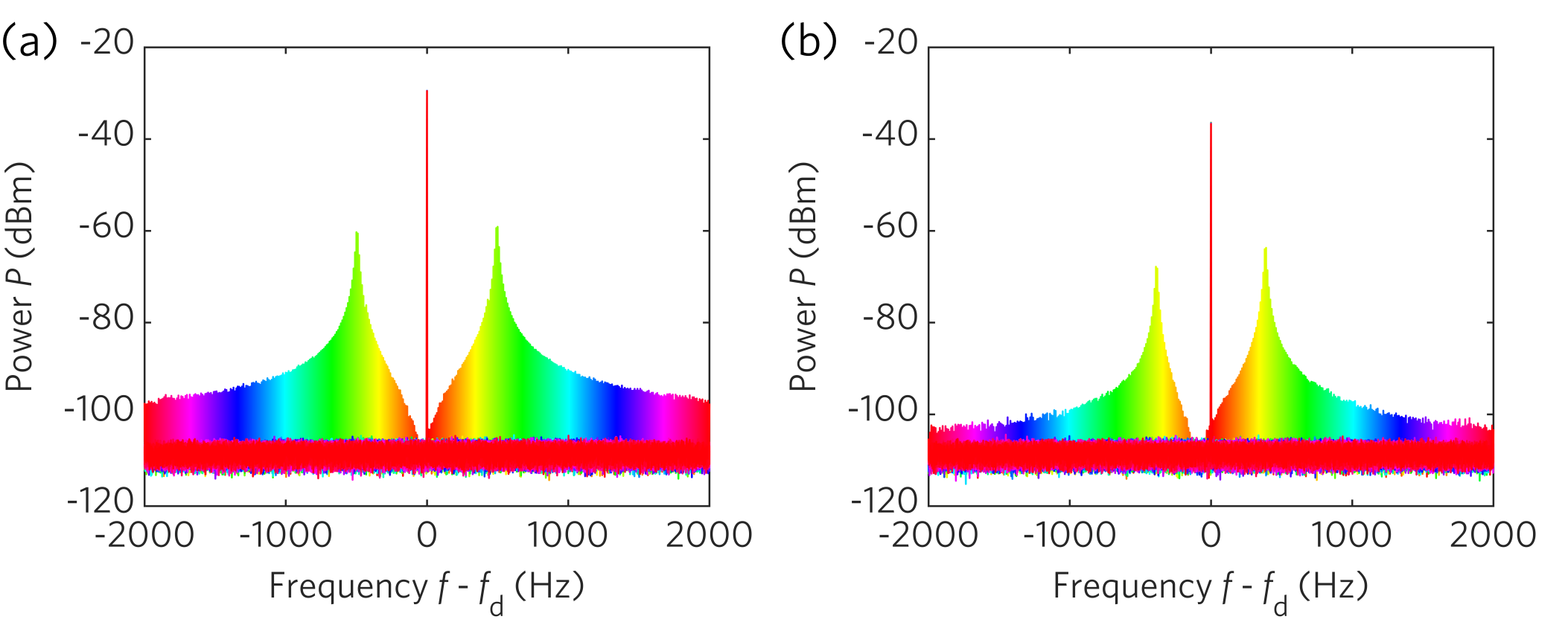}
    \caption{Power spectra of signal modulation in the nonlinear regime. a),b) Frequency mixing response with probe frequencies $f_\mathrm{p}$ increasing from 10 to 2000 Hz in steps of 10 Hz: the measured spectra are superimposed in one figure. The drive tone $V_\textrm{exc}$ = 0.5 V, probe tone $V_\mathrm{p}$ = 0.5 V, drive frequency $f_\mathrm{d}$ swept up to a) 255 kHz, b) 252 kHz, respectively. The color coding aims at mark the measured specific pair of the response at $f_\mathrm{d} \pm f_\mathrm{p}$.}
    \label{fig:SM_mod_response}
\end{figure*}

\begin{figure}[htbp]
 \includegraphics[width=0.9\linewidth]{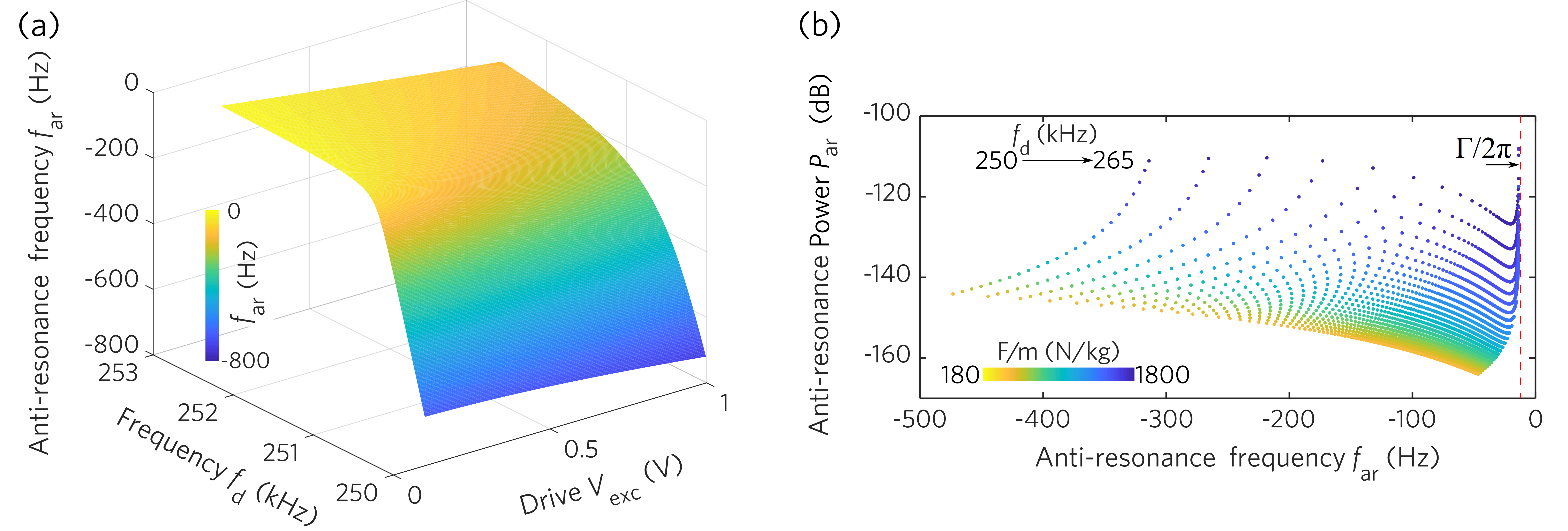}
     \caption{Theoretically calculated $M(f_\mathrm{ar},P_\mathrm{ar})$, the antiresonance frequency $f_\textrm{ar}$ and its corresponding power $P_\mathrm{ar}$. a) theoretically calculated $f_\textrm{ar}$ with different $f_\mathrm{d}$ and $V_\textrm{exc}$. b) Theoretically calculated coordinate map of $M$ vs. $f_\mathrm{d}$ (from 250 kHz up to 265 kHz) and $V_\textrm{exc}$ (converted to $F/m$, presented as color code). The red dashed line reveals the minimum value of $\omega_\mathrm{ar}/2\pi = \Gamma/2\pi$.}
 \label{fig:SM_M_map}
\end{figure}

\begin{figure}[htbp]
 \includegraphics[width=0.5\linewidth]{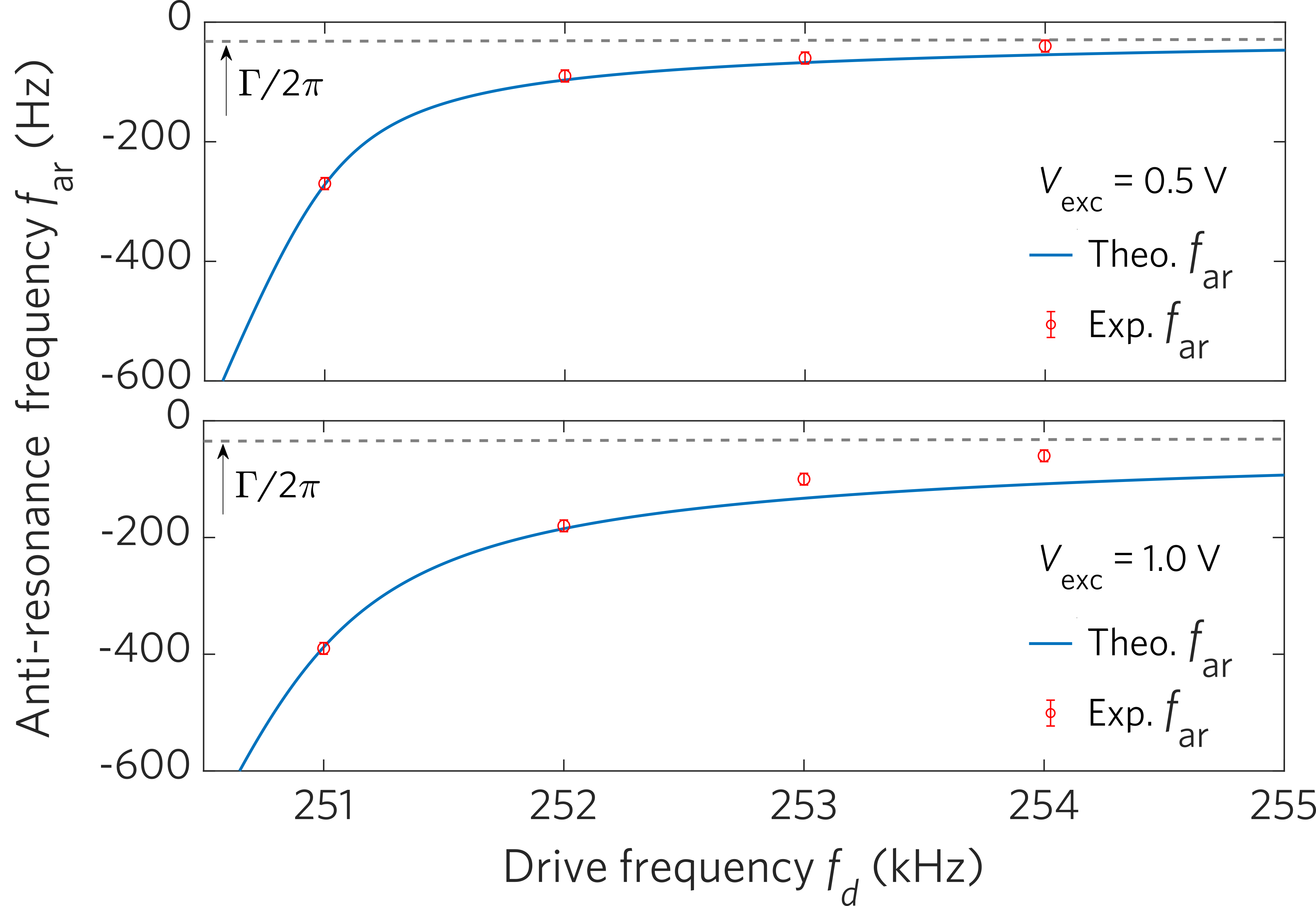}
     \caption{Theoretically calculated $f_\textrm{ar}$ while sweeping $f_\mathrm{d}$ at $V_\textrm{exc} = 0.5~\mathrm{V}$ (upper panel) and $V_\textrm{exc} = 1~\mathrm{V}$ (lower panel). The experimentally measured $f_\textrm{ar}$ under different $f_\mathrm{d}$ and corresponding $V_\textrm{exc}$ are superimposed in each panel with an error of $\pm$ 10 Hz. The grey dashed line reveals the minimum value of $\omega_\mathrm{ar}/2\pi = \Gamma/2\pi$.}
 \label{fig:SM_lincut_fdip}
\end{figure}

\indent We showed additional experimental results of the spectral response of sidebands under different drive frequency $f_\mathrm{d}$ and driving strength $V_\textrm{exc}$.
By comparing Fig. \ref{fig:SM_mod_response} a), b), and Fig. 2a) in the main text, different degrees of asymmetry between the blue and red sidebands can be observed. With decreasing detunings, the silent region becomes more apparent as it widens and shifts away from the drive frequency. \\
\indent The dependence between the drive and the antiresonance can be better explored using Eq.\eqref{eq:dip_freq}. Figure \ref{fig:SM_M_map}a) shows a more complete diagram between the antiresonance frequency $f_\textrm{ar}$ and different drive condition $f_\mathrm{d}$ and $V_\textrm{exc}$ for the (1,1) mode using Eq. (\ref{eq:dip_freq}). We can clearly see that $f_\textrm{ar}$ is large when the detuning $\delta \omega/2\pi$ is small and $F$ is large. 
In Fig. \ref{fig:SM_M_map} b), we present the theoretically calculated coordinate $M(f_\mathrm{ar},P_\mathrm{ar})$ in the sideband spectrum by changing $f_\mathrm{d}$ (marked in the figure) and $F/\textrm{m}$ (color bar). The minimum $f_\textrm{ar} = \Gamma/2\pi$ can be easily identified from Eq. (\ref{eq:dip_freq}), which is indicated by the red dashed line in Fig. \ref{fig:SM_M_map}b). 
A more detailed comparison for $f_\textrm{ar}$ between the theoretical and experimental results can be found in Fig. \ref{fig:SM_lincut_fdip}, where the drive condition are $V_\textrm{exc} = 0.5 ~\mathrm{V}$ and $1~\mathrm{V}$ in the upper and lower panel, respectively. The experiment data in Fig. \ref{fig:SM_lincut_fdip} (red circles) are the same with those presented in Fig. 3 of the main text (white dots). In both panels, the experimental data deviates from the theory as the detuning increases, suggesting the presence of other nonlinear effect such as the spatial modulation or mode coupling \cite{yang2019spatial, yang2021persistent}. 

\begin{figure*}[htbp]
    \includegraphics[width=\linewidth]{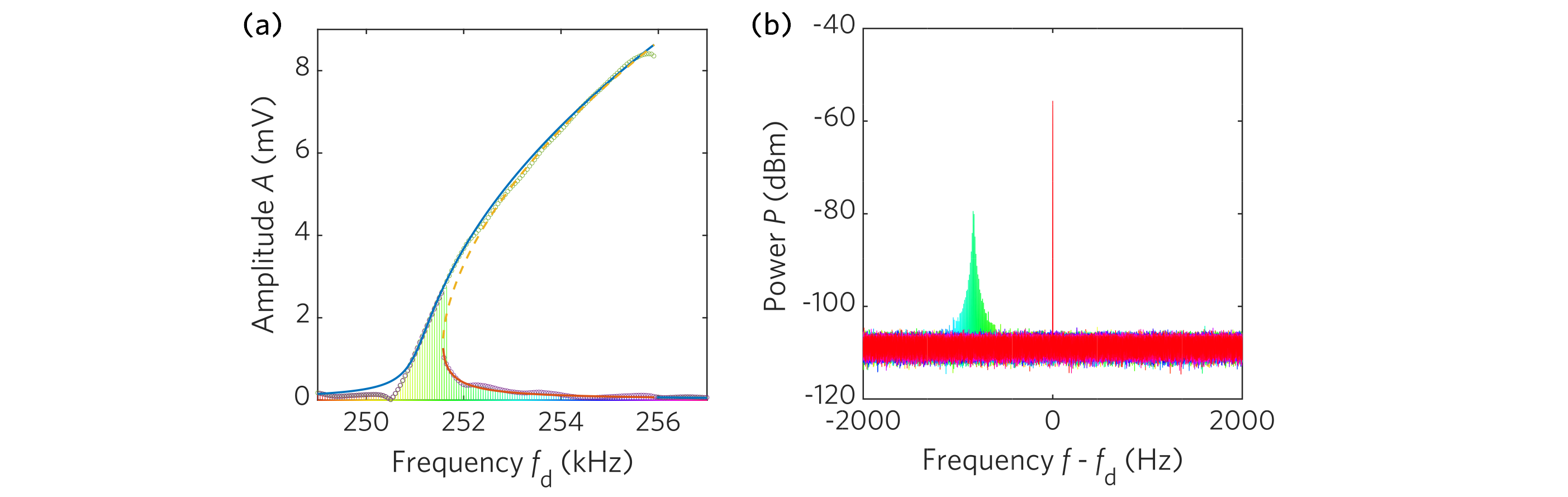}
    \caption{ a) For- and backward swept Duffing resonance of the (1,1) mode measured with $V_\textrm{exc}$ = 0.5 V, plotted as green and purple dots, respectively. Solid curves show theoretical results using the Duffing model. The dashed yellow line corresponds to the unstable solution of the Duffing model. The colored spikes area shows the power spectra of the vibration response, when switching on the drive at $f_\mathrm{d}$ (without sweeping from lower frequency) with amplitude $V_\textrm{exc}$ = 0.5 V. b) Power spectra when directly switching on the drive with $V_\textrm{exc}$ = 0.5 V at frequency $f_\mathrm{d}$ = 252 kHz and $f_\mathrm{p}$ increasing from 10 to 2000 Hz in steps of 10 Hz with $V_\mathrm{p}$ = 0.5 V; all measured spectra superimposed in one figure. }
    \label{fig:SM_low_branch}
\end{figure*}

\indent We wish to make two comments before ending this section. First, depending on whether the drive slowly sweeps from a negative detuning to $f_\mathrm{d}$ or directly jump to $f_\mathrm{d}$, the resonator could be at different stable states and hence different sideband response. As a demonstration, Fig. \ref{fig:SM_low_branch} b) showed the different vibrational state of the (1,1) mode with different sweeping methods, where $V_\textrm{exc}$ = 0.5 V and $f_\mathrm{d}$ ranges from 249 kHz to 257 kHz at a step size of 10 Hz. The green coloumns are the amplitudes from direct turn-on of the drive frequency whereas colored circles denote the for- and backward sweeping of the resonance curve. The blue solid line denotes the fitting using the Duffing model. As evident, directly turning on the drive only puts the resonator in the low-amplitude state if the drive frequency is within the bistable regime (e.g., $f_\mathrm{p}$ = 252 kHz). In this situation, the two-tone measurement ($f_\mathrm{p} \ll f_\mathrm{d}$) only excites the red sidebands, as shown in Fig. \ref{fig:SM_low_branch}c), drastically different from the sideband response for the high-amplitude states.\\
\indent Second, we emphasise that if $f_\mathrm{d}$ is set to far-off-resonance, only the two signals $f_\mathrm{d}$ and $f_\mathrm{p}$ are present with no evidence of any other tones. Spectral purity is crucial to the accuracy of the measurement.\\

\section{amplitude-frequency modulation}
\begin{figure*}[htbp]
    \includegraphics[width=\linewidth]{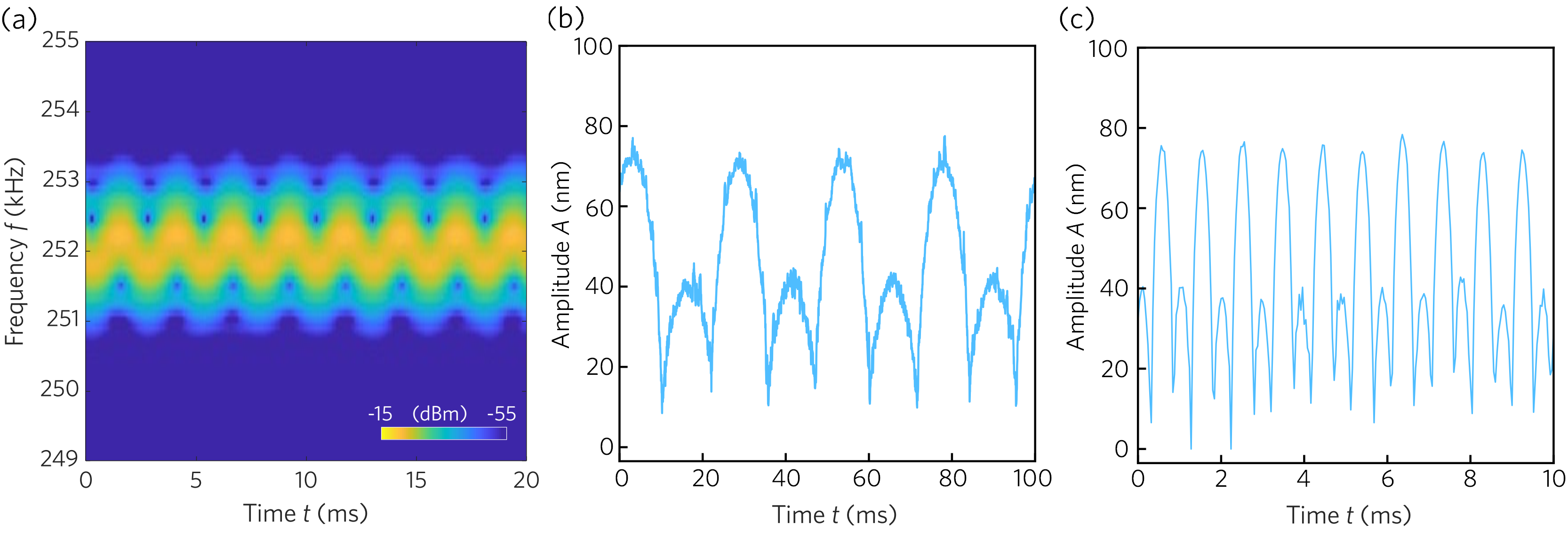}
    \caption{a) The power spectra of the signal modulation in the nonlinear regime, $V_\textrm{exc}$ = 0.5 V, $f_\mathrm{p}$ = 300 Hz and $f_\mathrm{d}$ = 252 kHz. b)-c) Measured amplitude modulation response by imaging white light interferometry (IWLI) using two-tone measurements. $V_\textrm{exc}$ = 1.0 V and the $V_\mathrm{p}$ = 0.5 V.  in b) and c) The $f_\mathrm{p}$ - $f_\mathrm{d}$ is set to 0.04 Hz and 1.0 Hz.
    }
    \label{fig:SM_freq_amp_mod}
\end{figure*}

\indent The sideband response in the frequency domain is also evidenced by its time-domain power spectrum with an amplitude frequency modulation \cite{pai2008detection, moore1992detection}, shown in Fig. \ref{fig:SM_freq_amp_mod}.  
In panel a), we show the time resolved frequency spectra obtained by fast fourier transformed (FFT) of the real-time oscillation trace with a drive tone of $f_\mathrm{d}$ = 252 kHz and $V_\textrm{exc}$ = 0.5 V and a probe tone of $f_\mathrm{p}$ = 380 Hz and $V_\mathrm{p}$ = 1 V using an oscilloscope. The freuqnecy modulation is self-evident from the periodically changing stiffness of the resonator and an asymmetry of the amplitude.\\
\indent The amplitude modulation can be better resolved using an imaging white light interferometer (IWLI). Detailed setup information is presented in our previous publications \cite{yang2017quantitative,yang2019spatial,yang2021persistent,waitz2012mode,waitz2015spatially}. The membrane resonator is first driven into a stable nonlinear vibration state of the (1,1) mode with $f_\mathrm{d}$ = 985 kHz and $V_\textrm{exc}$ = 1 V. The additional probe tone of $V_\mathrm{p}$ = 0.5V is added into the piezo actuator with the method described in Fig. 1d) at $f_\mathrm{p}$ = $f_\mathrm{d} - 0.04~\textrm{Hz}$ and  $f_\mathrm{p}$ = $f_\mathrm{d}- 1~\textrm{Hz}$. The IWLI monitors the vibrational amplitude over the entire resonator and gives a mean amplitude. 
As shown in Fig. \ref{fig:SM_freq_amp_mod}b)-c) for different $f_\mathrm{p}$, we clearly observe an amplitude modulation by noting a modulating amplitudes with a frequency of 0.04 Hz and 1 Hz under a fixed probe voltage $V_\mathrm{p}$ = 0.5 V. The resulting beating frequencies are equal to $f_\mathrm{p} - f_\mathrm{d}$. We note that the smaller half nodes are a result of the data processing of the IWLI which only gives the modulus of the deflection.\\ 

\section{Noise-induced sideband effects of a Duffing resonator}

\begin{figure*}[htbp]
    \includegraphics[width=\linewidth]{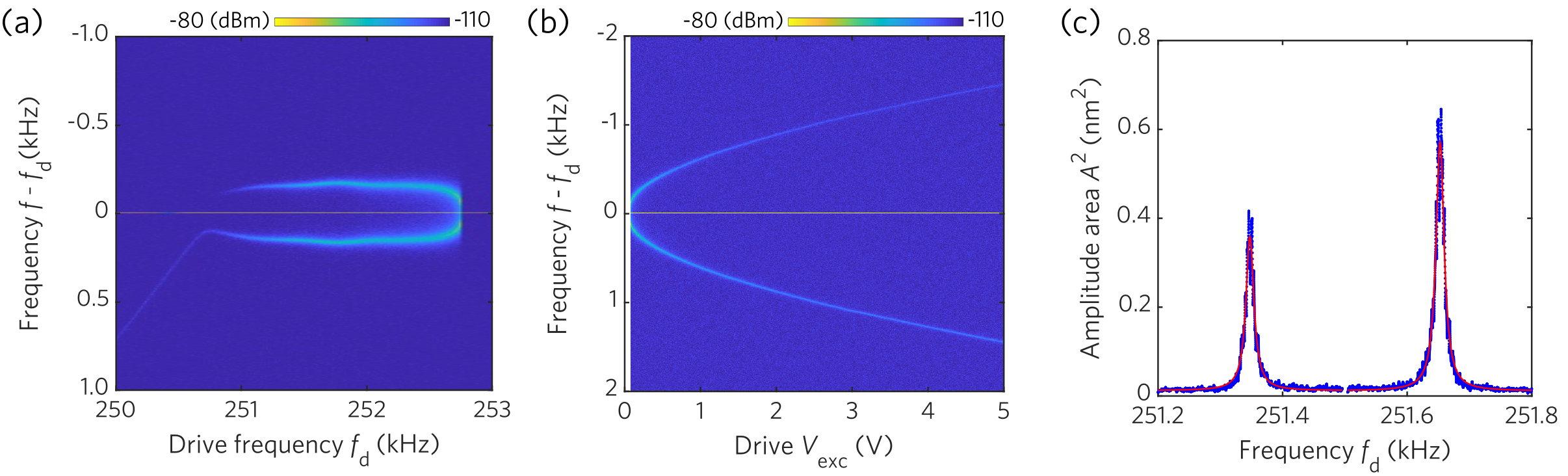}
    \caption{Measured detuning and drive-dependent noise-induced sideband power spectra in the Duffing regime. a) Detuning-dependent sidebands measured with $V_\textrm{exc}$ = 0.1 V. b) Drive-dependent sidebands measured at $f_\mathrm{d}$ = 252.5 kHz. c) Sideband peaks cut out from the power spectrum for $f_\mathrm{d}$ = 251.5 kHz from a), fitted by a Lorentzian function (red solid line).}
    \label{fig:SM_SB_measure}
\end{figure*}

In this section, we discuss the noise-induced sideband spectra. Fig.\ref{fig:SM_SB_measure} shows the sideband dependence of the (1,1) mode on detuning and drive power. The center bright line indicates the resonant response of the mode at the drive frequency. When the membrane resonator is driven at the negative detuning, only the blue sideband can be observed. In b) and c), For positive detuning, the sideband appears symmetrically with respect to $f_\mathrm{d}$. In addition, the area of the sideband becomes larger and the amplitudes of the red and blue sidebands gets stronger with increasing $f_\mathrm{d}$. These sidebands can be well fitted by a Lorentzian function, as shown in Fig. S8 c). \\

\indent The sideband responses described above are consistent with those reported in Ref. \onlinecite{huber2020spectral} where the sidebands can be excited by pure environmental thermal fluctuations thanks to its ultrahigh $Q$-factor. In our device, the $Q$ is one order of magnitude smaller than that in Ref. \onlinecite{huber2020spectral}. Nevertheless, we can still access these sideband by introducing an additional white noise into the system, alleviating the stringent $Q$ requirement for sideband measurement.
\bibliographystyle{apsrev4-1}
\bibliography{modulation_references}